\documentclass[twocolumn,prl,floatfix,epsfig]{revtex4}

\usepackage[dvips]{epsfig}
\usepackage{bm}
\usepackage{amsmath,amssymb}

\begin{document}

\title{Spherical Foams in Flat Space}
\author{Carl D. Modes and Randall D. Kamien}  \affiliation{Department of Physics and Astronomy, University of Pennsylvania, Philadelphia, PA 19104-6396, USA}

\date{\today}

\begin{abstract}
Regular tesselations of space are characterized through their Schl\"afli symbols $\{p,q,r\}$, where each cell has regular $p$-gonal sides, $q$ meeting at each vertex, and $r$ meeting on each edge.  {\sl Regular} tesselations with symbols $\{p,3,3\}$ all satisfy Plateau's laws for equilibrium foams.  For general $p$, however, these regular tesselations do not embed in Euclidean space, but require a uniform background curvature.  We study a class of regular foams on ${\bf S}^3$ which, through conformal, stereographic projection to $\mathbb{R}^3$ define irregular cells consistent with Plateau's laws.  We analytically characterize a broad classes of bulk foam bubbles, and extend and explain recent observations on foam structure and shape distribution.  Our approach also allows us to comment on foam stability by identifying a weak local maximum of $A^{\frac{3}{2}}/V$ at the maximally symmetric tetrahedral bubble that participates in T2 rearrangements.
\end{abstract}

\maketitle
From beer to bread, metals to mousse, and shaving cream to souffles, foams are space-and mouth-filling structures which arise in a broad array of applied contexts.  Interest in foams has grown \cite{JFM, Feitosa} as they have been used to model many other systems: from copolymers and fuzzy colloids \cite{Grason,Ziherl}, to the Edwards' picture of granular matter \cite{Edwards}, and even to space-time and cosmology \cite{Gott,Wheeler}.  Unfortunately, the details of a foam's bulk interior have long evaded both direct experimental probe \cite{Matzke} and exhaustive theoretical treatment \cite{PoF}.  At issue is the topological complexity of the network of foam bubble borders and the interaction of these borders with light, along with the disorder promoted by the incompatibility of Plateau's laws with space-filling regular polyhedra.  Recall that Plateau posited \cite{Plateau,Taylor} that three-dimensional dry foams would satisfy two rules: 1) each vertex terminates four edges, each an angle $\delta=\cos^{-1}(-\frac{1}{3})$ apart from the other, and 2) each edge is the border of three faces, and the dihedral angle between these faces is $2\pi/3$.    In this letter we study the properties of a large class of bubbles that extend the notion of ``isotropic Plateau polyhedra'' \cite{Hilg} or ``average $\cal N$-hedra'' \cite{Glicks} which we will call ``average polyhedra'' in the following.  We determine both the area $A$ and volume $V$ of these cells and find that specific area, $A^{\frac{3}{2}}/V$ is nearly constant for the class of dodecahedral and tetrahedral bubbles, with the exception of a weak local maximum around the symmetric tetrahedron.  Our more general approach is a new step toward understanding the statistics of random foams \cite{Kraynik}, and the landscape near the symmetric 4-sided bubble offers insight into foam evolution at finite temperature.  The projections that we make are all Plateau-law-abiding foams.

\begin{figure}[t!]
\centering
\epsfig{file=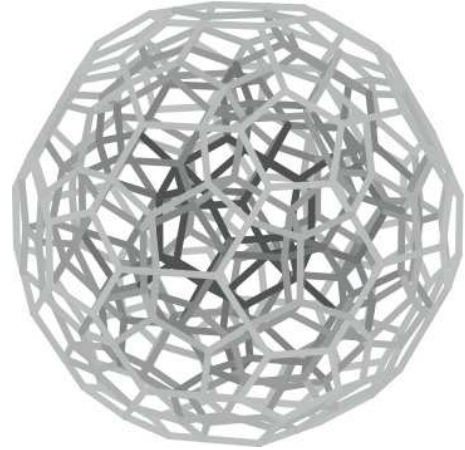,width=6cm}
\caption[Flat projection of the 120-cell]{A simple flat (though non-conformal) projection of the 120-cell into $\mathbf{R}^3$.  Darker lines are projected closer to the origin, lighter lines further.  Note the dodecahedral nature of the cells and their foam-like coordination. \label{120}}
\end{figure}  

The story in two dimensions is qualitatively different: there the decoration theorem \cite{decorate} and the von Neumann-Mullins growth law \cite{vNM} provide a coherent and useful description of foams.  In three-dimensions, geometrical frustration cuts our understanding at its core -- indeed, only recently was the growth law generalized above two dimensions \cite{MacPherson} and there are no regular tesselations of $\mathbb{R}^3$ which satisfy Plateau's laws.  Happily, it is well known that, in some cases, this type of frustration may be lifted by the introduction of a background curvature to the system \cite{blue, CDMRDK}.  In order to progress along these lines in the case of foams, it is expedient to consider Plateau's laws in a language that allows for a simple connection to the curvature.  That language is the geometric language of regular tesselations of space \cite{polytopes}.  To classify such tesselations we use the Schl\"afli symbols, $\{p,q,r\}$, where $p$ is the number of edges on each face of the cell,  $q$ is the number of these faces meeting at each vertex of \textit{one} cell, and $r$ is the number of cells or faces meeting around each edge.  By way of example, the only such tesselation allowed in flat space is the cubic honeycomb.  Since the cells are cubes,  $p=4$ and $q=3$; four cubes fit together around an edge, and hence the Schl\"afli symbol is $\{4,3,4\}$.  The cubic honeycomb does not satisfy Plateau's rules as four faces meet at an edge.  The requirement of dihedral angles of $2\pi/3$ implies that $r=3$.  Since each vertex must terminate four edges, it follows that each cell must have three such edges and thus $q=3$.  Thus we see that 
any regular space tesselation of the form $\{p,3,3\}$ necessarily represents a valid foam.  If we divide each $p$-gonal face into $p$ isoceles triangles, each will have one angle of $2\pi/p$ and two angles $\frac{\delta}{2}$.  On a uniformly curved space of constant sectional curvature $K$, this triangle, $\Delta$, lies in a two-manifold of curvature $K$.  The Gauss-Bonnet theorem relates the excess angle of the triangle to the integrated curvature:
\begin{equation}\label{eq:GB}
\frac{2\pi}{p} + 2\times\frac{1}{2}\cos^{-1}\left(-\frac{1}{3}\right) -\pi= \int_\Delta K dA
\end{equation}
In the flat limit $K=0$ and thus $p\approx 5.1043$; real foams are known to have $\sim \! 5.1$ sides per face on average \cite{Matzke, Kusner} as do the average polyhedra \cite{Glicks, Hilg}.
Of course, $p$ must be a positive integer and the theory of polytopes \cite{polytopes} limits us to $p\le 5$ (we shun the frumious horocycle -- $\{6,3,3\}$).  It follows that the regular polytopes can be embedded in a three-space with $K>0$, {\sl i.e.} the three-sphere, ${\bf S}^3$.  These ``polychora'' are the analogs of the two-dimensional tilings of ${\bf S}^2$ which correspond to the platonic solids.  
We choose to study two tesselations of $\mathbf{S}^3$,  the so-called 5-cell, $\{3,3,3\}$ and the 120-cell, $\{5,3,3\}$ (Figure \ref{120}).  The 120-cell is of interest because this regular polytope requires the least amount of background curvature to realize among the family in question.  Meanwhile, the 5-cell is worthwhile because it features tetrahedral bubbles like those seen at ``T2'' topological rearrangements, where a foam bubble shrinks until it disappears, leaving behind a four-fold vertex \cite{PoF}.   Related constructions have been used to show the uniqueness of $k$-bubble clusters in $d$ dimensions \cite{AJM}.

But how can we make contact with physical foams residing in actual, flat space?  We observe that the constraints of Plateau's laws that forced the form $\{p,3,3\}$ are all constraints on angles, hence mappings from the $3$-sphere to $\mathbb{R}^3$ that preserve angles will also preserve Plateau's laws.  We choose here to work with the conformal stereographic projection, long familiar to cartographers \cite{Ptolemy} where $X=x/(1-w), Y=y/(1-w)$, and $Z=z/(1-w)$ where $X,Y,Z$ are cartesian coordinates in the target $\mathbb{R}^3$, and $x,y,z,w$ are the coordinates in $\mathbf{S}^3\subset \mathbb{R}^4$ with $x^2+y^2+z^2+w^2=1$. 

\begin{figure}[t!]
\centering
\epsfig{file=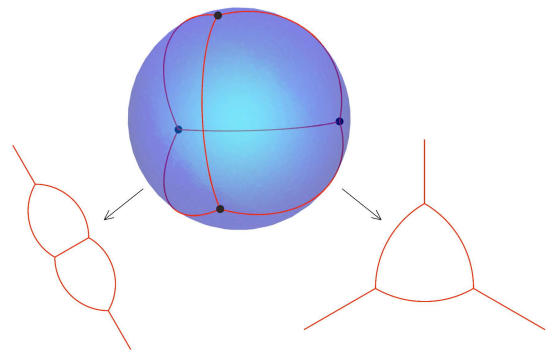,width=8cm}
\caption[Tetrahedral covering of the sphere projected to a Reuleaux triangle]{(color online).  The foam-like tetrahedral covering of the sphere and its Reuleaux Triangle image under stereographic projection from a vertex.  The full family of possible conformal distortions of this triangle is accessible by direct geometric calculation. \label{Reuleaux}}
\end{figure}
In two dimensions, this projection is the standard stereographic projection of ${\bf S}^2$ to $\mathbb{R}^2$ and we may project the $\{p,3\}$ Platonic solids (the tetrahedron, cube, and dodecahedron) onto the plane.  For concreteness, we will consider the $\{3,3\}$ tetrahedron shown in Fig. \ref{Reuleaux}; the conformal projection preserves the $2\pi/3$ angle between edges and so we will find a two-dimensional cell that satisfies Plateau's rules \cite{Wrecent}.  With one vertex at the pole, $(0,0,1)$, the remaining set of vertices is $\mathcal{T}=\{(\sqrt{\frac{2}{3}},-\frac{\sqrt{2}}{3},-\frac{1}{3}),(-\sqrt{\frac{2}{3}},-\frac{\sqrt{2}}{3},-\frac{1}{3}),(0,\frac{2\sqrt{2}}{3},-\frac{1}{3})\}$ and they are all connected by arcs of great circles.  We have the freedom to choose from which point we project ${\bf S}^2$, and each projection will give us two-dimensional cells that can fit in a two-dimensional foam.  If, for instance, we project through $(0,0,1)$, then the top vertex is projected to infinity and the remaining three vertices become $\mathcal{T}'=\{( \frac{\sqrt{3}}{2\sqrt{2}}, \frac{-1}{2\sqrt{2}}),(\frac{-\sqrt{3}}{2\sqrt{2}}, \frac{-1}{2\sqrt{2}}),(0, \frac{1}{\sqrt{2}})\}$.  Since great circles map to circles or lines, the edges of the
tetrahedron become arcs of circles and, by symmetry, we have the 
famous Reuleaux Triangle \cite{reuleaux}, familiar in two-dimensional foams as the ideal corner bubble during T2 topological rearrangements.  Were we to project from another point, or equivalently, rotate the sphere and project through the pole, we would find asymmetric foams.  Moreover, it is not just the central bubble that satisfies Plateau's laws; the entire projection is guaranteed to have all the correct angles for an equilibrium foam.  The central cell on which we focus could exist in any equilibrium $\mathbb{R}^3$ foam.

\begin{figure}[t!]
\centering
\epsfig{file=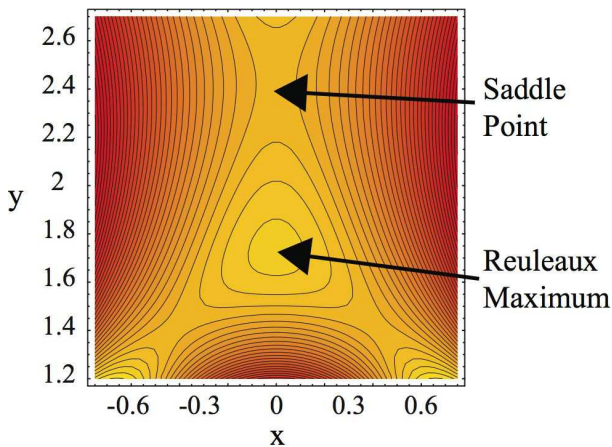, width=8cm}
\caption[Contour plot of $A^{\frac{3}{2}}/V$ as a function of the triangle coordinate $(x,y)$.  The Reuleaux triangle is a local maximum, surrounded by three deep valleys. ]{(color online).  Contour plot of $A^{\frac{3}{2}}/V$ as a function of the triangle coordinate $(x,y)$.  The Reuleaux triangle is a local maximum, surrounded by three deep valleys.    \label{localmax}}
\end{figure}

Working in $\mathbb{R}^2$, we can calculate both the perimeter and area of a general triangular bubble in a two-dimensional foam.  Here Plateau's rules are greatly simplified: edges must be arcs of circles and must meet at an angle of $2\pi/3$.   In order to parameterize an arbitrary triangle, we set two of the vertices at $(-1,0)$ and $(1,0)$, and allow the third to float as $(x,y)$, where $x$ and $y$ are both non-negative.   It is tedious but straightforward to calculate both the perimeter and the area as a function of $x$ and $y$.  The exact formulae are complicated and the results are better characterized graphically in Figure \ref{localmax}.

The machinery carries through in exactly the same way in the three dimensional case as well, allowing us to produce a continuous family of space foam bubbles.  In principle, this technique can produce bubble families from each of the $\{3,3,3\}$, the $\{4,3,3\}$, and the $\{5,3,3\}$ with 4-faced, 6-faced, and 12-faced bubbles, respectively.  The class of transformations that belong to the conformal group of $\mathbb{R}^3$ is restricted to solid body rotations and translations, reflections, dilations, and spherical inversions; the degrees of freedom available in the stereographic projection precisely account for each of these possibilities.  Generically, parallel translation of the hypersphere being projected relative to the target hyperplane yields translations.  Normal translation gives rise to dilations.  Rotations and reflections of the hypersphere about hyperplanes that lie in the target hyperplane give the solid body transformations.  The rest of the rotations give the spherical inversions.  Since the shape of a bubble or bubble cluster is sensitive only to inversions from among these possible transformation types, we can ignore the translational degrees of freedom and focus only on the out-of-hyperplane rotations.

Guided by experiment \cite{Matzke}, we choose first to examine the most prevalent from among these bubble types, those with twelve faces. 
As in the two-dimensional example, we will have the freedom to choose the coordinate orientation of the 120-cell inside ${\bf S}^3$ and the ability to choose the base point of the stereographic projection; either one of these choices completely exhausts the degrees of freedom available, leaving the other to parameterize the resulting family of bubbles.  We choose to pin the base point of the projection to the north pole of $\mathbf{S}^3$, $(0,0,0,1)$, and will rotate the 120-cell inside $\mathbf{S}^3$ to generate different bubbles.  In so doing it becomes transparent to identify the parameterization of the foam bubble family with the solid body rotation taking place on the 120-cell pre-image.  Said another way, we may unambiguously index our final bulk foam bubbles by elements of $\mathbf{SO(4)}$.  Note that, due to the high symmetry of the pre-image 120-cell, it will suffice to consider the action of the projection on only one cell, since one cell may be carried into the position of any other by some element of $\mathbf{SO(4)}$.  In analogy to the two-dimensional case, planar regions in ${\bf S}^3$ map to spherical caps in $\mathbb{R}^3$.  

If we view our parameter space as a higher dimensional rotation space, we will need to specify an unrotated, reference state.  We choose this state to coincide with the highest symmetry target bubble, the average polyhedra \cite{Hilg,Glicks}, whose vertices coincide with the vertices of a regular dodecahedron and whose faces are all equivalent pentagonal spherical caps.  This high-symmetry target bubble may be obtained by orienting the pre-image 120-cell so that its vertices lie at $\frac{1}{2\sqrt{2}}$ of the permutations of \cite{polytopes} 
$(\pm 2, \pm 2, 0, 0)$,
$(\pm \sqrt{5}, \pm 1, \pm 1, \pm 1)$,
$(\pm \phi, \pm \phi, \pm \phi, \pm \phi^{-2})$, and
$(\pm \phi^{2}, \pm \phi^{-1}, \pm \phi^{-1}, \pm \phi^{-1})$ 
and $\frac{1}{2\sqrt{2}}$ of the even permutations of 
$(\pm \phi^{2}, \pm 1, \pm \phi^{-2}, 0)$,
$(\pm \sqrt{5}, \pm \phi, \pm \phi^{-1}, 0)$, and
$(\pm 2, \pm \phi, \pm 1, \pm \phi^{-1})$
where $\phi = (1+\sqrt{5})/2$ is the Golden mean.

With this characterization of the reference 120-cell in hand, we may begin to apply elements of $\mathbf{SO(4)}$ and project down to flat space.  In analogy with the example calculation shown above for the projection of the tetrahedron from $\mathbf{S}^2$ we obtain the equations of the spheres in $\mathbb{R}^3$ defining the faces of the image dodecahedral bubble.  Qualitatively, rotating the pre-image 120-cell around one of the $w$ planes results in a bubble that has been stretched, with a vertex or edge pulled farther away from the others.  This stretching also controls the sign of the face curvature for the bubbles; we consider relatively small deformations in order to preserve the positivity of the curvature for all twelve faces.

\begin{figure}[t!]
\centering
\epsfig{file=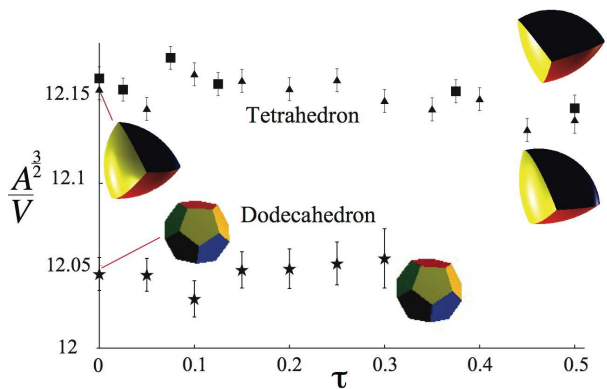,width=8cm}
\caption[Dimensionless surface area to volume ratios for dodecahedral and tetrahedral bubbles distorted by rotations in $w$]{(color online). Dimensionless surface area to volume ratios for dodecahedral (lower curve) and tetrahedral bubbles (upper curve) distorted by rotations in the $w$-$z$ plane by $\tau$ followed by sterographic projection.  Two different rotational starting points are used for the tetrahedra, with no significant difference in $A^{\frac{3}{2}}/V$.  Note the near independence of both bubble types' ratios on the mapping parameter, $\tau$, with the exception of the maximum and valley near $\tau = 0$ for the tetrahedron.  Likewise, the distorted shapes are still highly constrained by Plateau's laws.  For a sphere, $A^{\frac{3}{2}}/V=6\sqrt{\pi} \approx 10.635$ attains its minimum.  \label{flatness}}
\end{figure}

Taking the appropriate face and cell intersections, we numerically determined the volume and surface area of the projected bubble.  Perturbing away from the maximally symmetric bubble by rotating around the $w$-$z$ plane by an angle $\tau$, we calculate the non-dimensionalized combination $A^{\frac{3}{2}}/V$.  One would expect this quantity to dominate the bubble's free energy and, through the Boltzmann factor, control the likelihood of finding a bubble of this shape in an arbitrary dry foam.  Interestingly, however, we find that to within the errors of our calculation $A^{\frac{3}{2}}/V$ is essentially constant with respect to $\tau$ (see Figure \ref{flatness}).  This result extends the observations of Hilgenfeldt {\sl et al.} who found that for their idealized Isotropic Plateau Polyhedra the non-dimensional surface area is largely insensitive to the number of faces \cite{Hilg}.  In addition, the demonstration that even ``stretching" foam bubbles does not change their reduced surface area-to-volume ratio serves to support and strengthen the recent results of Kraynik {\sl et al.}, who found in simulations using Surface Evolver that the average reduced surface area-to-volume ratio of native bulk foam bubbles is constant \cite{Kraynik}.  The implication, then, is that the bubble shape distribution in a bulk foam is dominated by the topology of the network of Plateau borders, with little contribution from bubble geometry.

Pursuing the tetrahedral target bubbles generated by $\{3,3,3\}$ is qualitatively similar to what was done with the 120-cell.  The unrotated, reference state vertices for the 5-cell in $\mathbf{S}^3$ are :
$\left(0, 0, 0, 1\right)$,$\left(0, 0, \frac{\sqrt{15}}{4}, -\frac{1}{4} \right)$,
$\left(0, \sqrt{\frac{5}{6}}, -\frac{1}{4} \sqrt{\frac{5}{3}}, -\frac{1}{4} \right)$,
$\left(\frac{1}{2} \sqrt{\frac{5}{2}}, -\frac{1}{2} \sqrt{\frac{5}{6}}, -\frac{1}{4} \sqrt{\frac{5}{3}}, -\frac{1}{4} \right)$, and
$\left(-\frac{1}{2} \sqrt{\frac{5}{2}}, -\frac{1}{2} \sqrt{\frac{5}{6}}, -\frac{1}{4} \sqrt{\frac{5}{3}}, -\frac{1}{4} \right)$.
Proceeding as before, we perturbed the maximally symmetric tetrahedral bubble by rotating an angle $\tau$ about the $w$-$z$ plane.  As is the case for the dodecahedral bubbles, we find a largely flat dependence of $A^{\frac{3}{2}}/V$ on $\tau$ (Figure \ref{flatness}).  Furthermore, in agreement with past work \cite{Kraynik}, we find that the values of the reduced surface area-to-volume ratio for the tetrahedral bubbles are essentially indistinguishable from those of the dodecahedral bubbles, differing by less than one percent on average.  It is noteworthy, though, that the tetrahedra are uniformly higher in $A^{\frac{3}{2}}/V$, consistent with the expectation that the closer a cell is to achieving 13.7 faces, the more efficiently it will enclose its volume.

Our numerical results in three dimensions also suggest there is a local maximum of $A^{\frac{3}{2}}/V$ at the maximally symmetric Reuleaux tetrahedron.  Whether the existence of this local maximum is an impediment to T2 transitions at finite temperature is a natural question to pursue. Since this feature is preserved in two dimensions, where $P^2/A$ as shown in Figure \ref{localmax}, it may be possible to do both experiment and simulation to address this issue.

We have generated a continuous class of dodecahedral and tetrahedral bubbles via conformal projection of known polychora and have shown that the specific area of these bubbles is roughly constant over our parameter space.  Our methods have also allowed us to uncover  a possible local maximum of $A^{\frac{3}{2}}/V$ at the T2-event bubbles in both two and three dimensions.  It remains an interesting open question as to whether this approach can be generalized to study $\{p,3,3\}$
polytopes for non-integer $p$ along the lines of the average polyhedra \cite{Glicks,Hilg}. 

It is a pleasure to acknowledge discussions with B.G. Chen, D.J. Durian, O.L. Halt, J. M. Kikkawa, R.B. Kusner, E. Matsumoto, and V. Vitelli.  This work was supported by NSF Grant DMR05-47320, a gift from L. J. Bernstein, and a gift from H. H. Coburn.

\end{document}